\documentclass[journal]{IEEEtran}
\usepackage{cite}

\usepackage{url}

\usepackage[pdftex]{graphicx}

\usepackage{amsmath}
\usepackage{amssymb}

\usepackage{array}

\begin{document}

\title{Relative Transfer Function Inverse Regression \\ from Low Dimensional Manifold}

\author{Ziteng Wang, Emmanuel Vincent, Yonghong Yan}

%



\maketitle

\begin{abstract}
In room acoustic environments, the Relative Transfer Functions (RTFs) are controlled by few underlying modes of variability. Accordingly, they are confined to a low-dimensional manifold. In this letter, we investigate a RTF inverse regression problem, the task of which is to generate the high-dimensional responses from their low-dimensional representations. The problem is addressed from a pure data-driven perspective and a supervised Deep Neural Network (DNN) model is applied to learn a mapping from the source-receiver \emph{poses} (positions and orientations) to the frequency domain RTF vectors. The experiments show promising results: the model achieves lower prediction error of the RTF than the free field assumption. However, it fails to compete with the linear interpolation technique in small sampling distances.
\end{abstract}


%
\IEEEpeerreviewmaketitle


\section{Introduction}


\IEEEPARstart{T}{he} acoustic properties of a room environment can be fully characterized by the collection of the Acoustic Impulse Responses (AIRs). An AIR relates two arbitrary poses inside the room: one for the source and the other for the receiver. In the multichannel setup, where the receiver includes multiple microphones, the Relative Transfer Function (RTF)~\cite{gannot2001signal} is more often used. The estimation of the RTF is essential in many applications, such as sound field reproduction~\cite{betlehem2005theory}, dereverberation~\cite{lin2007blind}, source localization~\cite{li2015estimation} and source separation~\cite{koldovsky2015spatial}. Over the years, the RTF estimation is based merely on the measured signals~\cite{cohen2004relative,talmon2009relative,taseska2015relative}. While the prior knowledge of the RTF given the source-receiver pose could provide additional performance benefits~\cite{talmon2013relative}, it remains less studied.

The RTF involves two AIRs, the knowledge of which usually relies on explicit physical models. For instance, the classical image-source method~\cite{allen1979image} is widely used for simulating small room acoustics. The method assumes the response to be contributed by multiple virtual image sources, and the simulation requires the room to be rectangular and the wall reflection coefficients known. If the room geometry can be estimated in practice, an approximation of the AIR can be obtained accordingly~\cite{asaei2012computational,asaei2014structured}. Considering a predefined pose space only, the AIRs can be alternatively parameterized based on the harmonic solution to the wave equation~\cite{samarasinghe2015efficient}. Moreover, in stationary rooms, AIR measurements can be collected in advance and help the modeling. With a limited number of measurements, compressed sensing with sparsity is introduced to interpolate AIR for the early parts~\cite{mignot2013room} or in the low frequency bins~\cite{mignot2014low}. When dense sampling is achievable, the simple data-based linear interpolation turns out to be an effective approach~\cite{nishino1999interpolating,vincent2013second}.


In some common scenarios, e.g., in conference rooms or cars, the stationary assumption made above is reasonable since the layout does not often change over time. Indeed, the AIRs in this case are confined to a low-dimensional manifold because the source-receiver pose now works as the only varying degree of freedom. This geometrical property was revealed by the manifold learning paradigm~\cite{talmon2012parametrization} that was first introduced to parameterize linear systems. It was then adopted to RTF modeling~\cite{laufer2013relative,laufer2015study} and applied in supervised~\cite{talmon2011supervised} and semi-supervised source localization tasks~\cite{laufer2016mr,laufer2016semi}, which was to associate the manifold with the source poses. The RTF encodes the Interchannel Level Difference (ILD) in decibels and the Interchannel Phase Difference (IPD) in radians. The low-dimensional structure of the ILD and IPD was shown in a binaural setup~\cite{deleforge20122d}. Based on a local linearity assumption on the manifold, the method of Probabilistic Piecewise Affine Mapping (PPAM) was proposed for a 2D sound localization task. Specially, PPAM learned a bijective mapping between the poses and the ILD and IPD, whereas only the interaural-to-pose regression was discussed for localization. The bijective mapping has also been generalized to the cases of multiple sources~\cite{deleforge2015acoustic}, co-localization of audio sources in images~\cite{deleforge2015co} and partially latent response variables~\cite{deleforge2015high}.

In this letter, we raise the problem of RTF inverse regression, which is defined as approximating the high-dimensional responses from their low-dimensional representations. The question is: how to acquire prior knowledge of the RTF given the source-receiver pose? As deriving an explicit physical model could not always be possible but representative examples of the acoustic environment can be collected in advance, especially with modern smart devices, the problem is addressed from a pure data-driven perspective. A Deep Neural Network (DNN) model is proposed to directly generate the frequency domain RTF vector given the source-receiver pose. The DNN model is advantageous in that it learns a globally nonlinear mapping from the data examples while preserving a local linearity, which matches the manifold structure implicitly. In the experiments, the acoustic space is sampled uniformly and the DNN model is tested in unseen poses. The evaluation is based on an absolute prediction error measure, while applying the generated RTFs to specific applications is not in the scope here.

The rest of this letter is organized as follows. The relevant definitions are given in Section II. The RTF inverse regression task and the possible solutions including the supervised DNN model are explained in Section III. The experimental setups and results are presented in Section IV and conclusions are drawn in Section V.


\section{Definitions}

In a reverberant room environment, the RTFs represent the coupling between a pair of microphones in response to the source signal. Denote the source as $s$ and the two AIRs from the source to the microphones as $h_1,h_2$, the observations at time $t$ are written as
\begin{equation}\label{eq1:sig}
  a_m(t) = h_m(t)*s(t) + v_m(t) ~~~~m=1,2
\end{equation}
where $*$ denotes convolution and $v_m$ is the noise component in the $m$th microphone. Under the narrowband approximation, the signals in the frequency domain are given by
\begin{equation}\label{eq1:sigfreq}
  A_m(l,f) = H_m(f)S(l,f) + V_m(l,f)
\end{equation}
where $l$ is the frame index, $f$ is the frequency index, $A_m,S$ and $V_m$ are the Short Time Fourier Transform (STFT) coefficients of $a_m, s$ and $v_m$, respectively, and $H_m$ is the Fourier transform of $h_m$. The RTF is defined as
\begin{equation}\label{eq:rtf}
  H(f)=\frac{H_2(f)}{H_1(f)}.
\end{equation}

The RTFs are known to be governed by these parameters: the size and geometry of the room, the reflection coefficients of the walls, and the source-receiver poses. Assuming the room characteristics to be stationary over time, the RTFs are thus confined to a low-dimensional manifold that can be associated with the source-receiver poses as
\begin{equation}\label{eq:rtfpose}
    H(f) = g(\Theta_s,\Theta_r)
\end{equation}
where $g:\mathbb{R}^{L}\rightarrow \mathbb{R}^{D}$ is defined as the mapping function and $\Theta_s,\Theta_r$ are the pose parameters in the intrinsic Cartesian coordinate system on the embedded manifold. The pose parameters include the position coordinates ($x, y, z$) and the direction variables (azimuth angle, elevation angle and rotation angle).

In anechoic conditions, the translation from the source-receiver pose to the RTF is straightforward under the free field assumption, which is given by the direct sounds with (\ref{eq:rtf}) and
\begin{equation}\label{eq:green}
  H_m(f)=\frac{\text{exp}(j\cdot 2\pi f ||\Theta_{r_m} - \Theta_s||_2 / c)}{4\pi ||\Theta_{r_m} - \Theta_s||_2}
\end{equation}
where $j$ is the complex unit, $||\cdot||_2$ denotes the Euclidian norm and $c$ denotes the speed of sound~\cite{samarasinghe2015efficient}. However, the association becomes complex in reverberant conditions.

\section{RTF inverse regression}

The task of RTF inverse regression is to learn the mapping function $g$ from a given set of pairwise examples $\{\mathcal{X}:\Theta_s,\Theta_r;~\mathcal{Y}:H(f)\}$, while no physical constraint is involved. Three possible solutions are discussed in the following.

Linear interpolation is the intuitive way. Provided that $\mathcal{X},\mathcal{Y}$ have the same geometric structure in their separate space, the response of a new pose $\widehat{\mathcal{X}}$ can be estimated as
\begin{equation}\label{eq:linearinter}
  \widehat{\mathcal{Y}}=\frac{1}{\sum \alpha_i}\sum_{i=1}^{I}\alpha_i \mathcal{Y}_i
\end{equation}
where $\mathcal{Y}_i$ is the response of the neighboring pose $\mathcal{X}_i$ and $\alpha_i$ is the weighting parameter correlated to the spatial distance from $\mathcal{X}_i$ to $\widehat{\mathcal{X}}$.

Although PPAM was not proposed for this task, an instantiation of $g$ was realized based on a piecewise linear approximation~\cite{deleforge2015acoustic}. The acoustic space is divided into $K$ local regions and each is characterized by the affine transform:
\begin{equation}\label{eq:ppam}
  \mathcal{Y}=\sum_{k=1}^{K}\mathbb{I}(k)({\bf A}_k\mathcal{X}+{\bf b}_k)+\epsilon
\end{equation}
where $\mathbb{I}(k)=1$ if $\mathcal{X}$ lies in the $k$th local region, and 0 otherwise. ${\bf A}_k\subset\mathbb{R}^{L\times D}$ is the weight matrix, ${\bf b_k}$ is the bias vector and $\epsilon$ is the error term described by the Gaussian distribution.

DNNs feature multiple hidden layers trained through error backpropagation. Theoretically, the expression in~(\ref{eq:ppam}) can be approximated by a neural network with one hidden layer~\cite{pinkus1999approximation}, which can be described as
\begin{equation}\label{eq:dnn}
  \mathcal{Y}={\bf W}^{(1)}\zeta({\bf W}^{(0)}\mathcal{X}+{\bf b}^{(0)})+{\bf b}^{(1)}
\end{equation}
where ${\bf W}^{(i)}, {\bf b}^{(i)}$ are the $i$th layer parameters and $\zeta(\cdot)$ is a nonlinear activation function. More details about the proposed DNN model are given as below.

\subsection{Inputs and targets}

The inputs of the DNN model are the pose parameters $\{\Theta_s,\Theta_r\}\subset\mathbb{R}^{L}$, the effective dimension of which is dependent on the degree of freedom in the system, since fixed input values should make no difference to the performance.

The target RTFs are complex vectors, which are represented by the real-valued ILD and IPD as
\begin{IEEEeqnarray}{rCl}\label{eq:ilpd}
  \text{ILD} &=& 20{\rm log}_{10}|H(f)| \\
  \text{IPD} &=& {\rm arg}(H(f)).
\end{IEEEeqnarray}
The sine and cosine of the IPD are computed and concatenated with the ILD to form the final target vector for training, which is of $D = 513\times3$ dimensions as STFT is performed in $1024$ points. The setup follows that in~\cite{deleforge2015acoustic}, nevertheless, alternative targets could be the real and imaginary parts of the RTF.

Since data normalization is known to help the training, a normalized target RTF is considered as
\begin{equation}\label{eq:dn}
  \overline{H(f)}=\frac{H(f)}{H_d(f)}
\end{equation}
where $H_d(f)$ is calculated using~(\ref{eq:rtf}) and (\ref{eq:green}). This could provide marginal benefit in the latter experiments.

\subsection{DNN Architecture and training}

The DNN model is a basic feed-forward neural network with all the layers linearly connected. The ReLU activation function is used for the hidden layers and linear activation is used for the output layer. Given the previous target selection, a local normalization is enforced to the IPD output nodes:
\begin{equation}\label{eq:ln}
  o_{f,-}=\frac{o_{f,-}}{\sqrt{o_{f,s}^2+o_{f,c}^2}}
\end{equation}
which means that the sum of squares of the sine part $o_{f,s}$ and the cosine part $o_{f,c}$ in the $f$th frequency bin should equal to one.

In the training stage, the layer weights are initialized with Gaussian samples (zero mean and deviation $\sqrt{1/\text{in\_size}}$) and the bias vectors are initialized with zeros. The model is optimized under the Mean Squared Error (MSE) loss criterion. The Adam method~\cite{kingma2014adam} is used to update the model parameters and the learning rate is adjusted adaptively. Layer normalization~\cite{ba2016layer} is applied to the hidden layers to speed up the model convergence. Other regularization techniques such as dropout and batch normalization, are not found helpful here.

\subsection{Evaluation metric}

As far as we know, there is no established measure to evaluate the approximation error of a high-dimensional vector. A mean absolute error metric is chosen here to straightly show the performance in each frequency:
\begin{equation}\label{eq:err}
  \mu_f = \frac{1}{N}\sum_{n=1}^{N}||\widehat{\mathcal{Y}}_n(f)-\mathcal{Y}_n(f)||_1
\end{equation}
where $\widehat{\mathcal{Y}}_n$ is the predicted response of the $n$th test sample. The 95\% Confidence Interval (CI) of $\mu_f$ is also calculated to give an idea of the error distribution. The ILD and IPD prediction errors are treated separately.

\section{Experiments and results}

\subsection{Experimental setup validation}

In the first place, we validate the experimental setup before reporting further results on simulated data. The CAMIL dataset\footnote{\url{https://team.inria.fr/perception/the-camil-dataset/}} is considered for this task as it includes real-world recordings labeled with true poses. A simulated dataset is first created following its original setup and the models are then tested on both the real and simulated data. We consider the experimental setup is validated if similar trends are observed in the performance. Note that the CAMIL dataset consists of binaural recordings that involves Head-Related Transfer Functions (HRTFs) and changes the definition of RTF inverse regression slightly. Nevertheless, this dataset is to our knowledge the only qualified public dataset, and the previous analysis can be generalized to this case.

\begin{figure}
\centering
\includegraphics[width=0.89\linewidth]{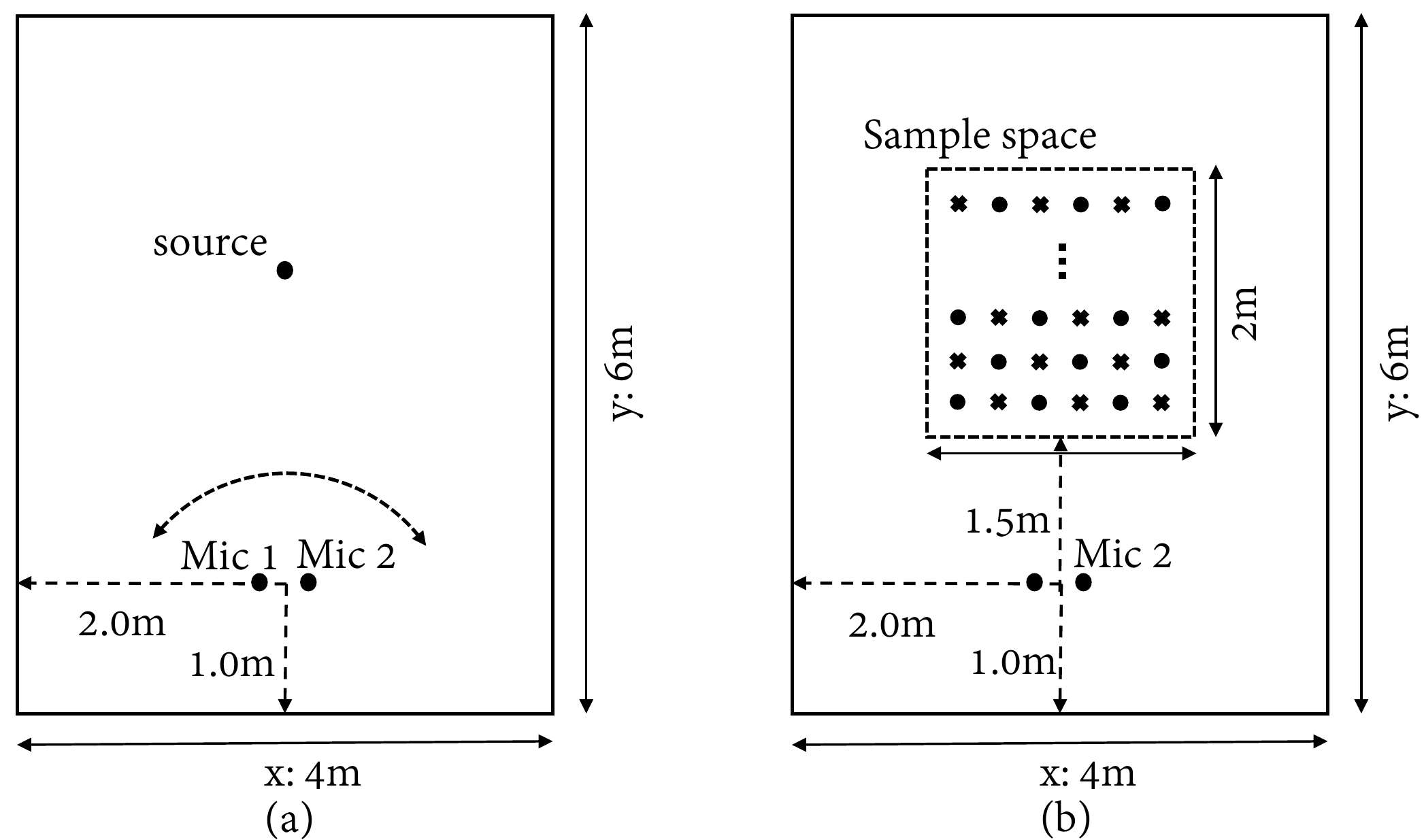}
\caption{Top view of the simulation setup: (a) simulated CAMIL dataset, (b) uniform acoustic space sampling. One every two samples (marked as dots in the sample space 2$\times$2$\times$1 m) is used in the training.}
\label{fig1:room}
\end{figure}

The simulation setup is illustrated in Fig.~\ref{fig1:room}(a). The room size is 4$\times$6$\times$3 m and the reverberation time is set to be 300 ms. A pair of microphones with cardioid directivity are used to imitate a dummy-head, while the AIRs (simulated using the image-source method~\cite{allen1979image}) are convolved with real measured  HRTFs\footnote{\url{https://dev.qu.tu-berlin.de/projects/measurements}}. The receiver is positioned at $[$2, 1, 1.4$]$, with microphone distance 0.18 m, and the source is fixed at 3 meters away in the front. The receiver pose is defined by an azimuth angle in the range of $[-160^{\circ}, 160^{\circ}]$ and an elevation angle in the range of $[-60^{\circ}, 60^{\circ}]$. Data samples are generated every $2^{\circ}$ and there are 9600 poses in total. For each pose, a one-second white noise signal is emitted from the source and then the captured microphone signals are used to calculate the corresponding RTF~\cite{deleforge2015acoustic}. This setup is one way to measure the RTF in practise, though it can be directly computed from the two AIRs with~(\ref{eq:rtf}) in the simulation.

For the DNN model, 3 hidden layers each with 1024 nodes are used throughout based on heuristic search. One every two samples in the dataset is used for training. Half of the left is used for development and the other half for testing. Early stopping is applied to avoid overfitting when the development set error no longer decreases after a patience of 5 epoches. The method of PPAM~\cite{deleforge2015acoustic} (source code provided therein) is investigated and the parameter $K$ is chosen from $\{$64, 128, 256$\}$ to achieve the best performance. It is worth to note again that PPAM was not proposed for the task here. Linear interpolation is also considered, but in a simplified way:
\begin{equation}\label{eq:li}
  \widehat{\mathcal{Y}}_n(f)=[\mathcal{Y}_{n,1}(f)+\mathcal{Y}_{n,2}(f)]/2
\end{equation}
where $\mathcal{Y}_{n,1}$ and $\mathcal{Y}_{n,2}$ are the response vectors of the spatially adjacent poses. The interpolated IPD is normalized as in~(\ref{eq:ln}).

The results on the real and simulated CAMIL datasets are given in Table~\ref{table:camil}. The mean values $\mu_f$ are further averaged over the frequencies. Similar trends are observed: DNN achieves lower prediction errors than PPAM in ILD but slightly higher errors in IPD, and both methods fail to compete with linear interpolation. This should validate our simulation setup. The reason that DNN and PPAM perform relatively better on the simulated data could be due to the simplification of the simulation setup.

\begin{table}[!h]
\caption{ILD/IPD prediction errors on the real (left part) and simulated (right part) CAMIL datasets.(mean$\pm$CI bound)}
\label{table:camil}
\begin{center}
\begin{tabular}{|l|c|c||c|c|}
  \hline
                     & ILD & IPD               & ILD & IPD \\ \hline \hline
  PPAM    & 1.73$\pm$.057 & 0.33$\pm$.015   & 1.42$\pm$.066 & 0.27$\pm$.013 \\ \hline
  DNN     & 1.42$\pm$.047 & 0.34$\pm$.014   & 1.23$\pm$.042 & 0.29$\pm$.013\\ \hline
  Linear  & 1.03$\pm$.035 & 0.18$\pm$.011	& 1.07$\pm$.034 & 0.20$\pm$.011\\ \hline

\end{tabular}
\end{center}
\end{table}

\begin{figure}
\centering
\includegraphics[width=0.94\linewidth]{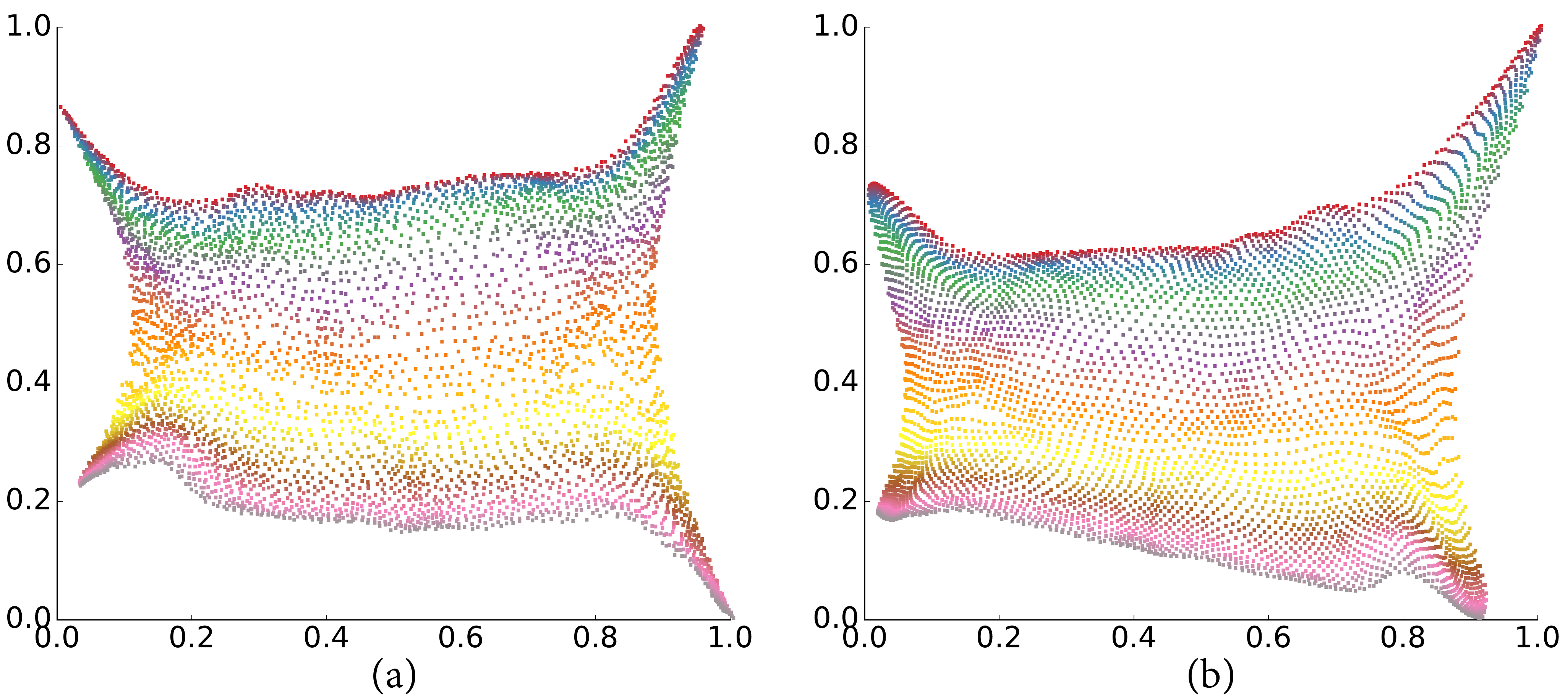}
\caption{2D visualization of the manifold using local tangent space alignment in the scikit-learn toolbox: (a) target ILDs, (b) DNN generated ILDs. Samples with the same elevation angle have the same color.}
\label{fig2:embed}
\end{figure}

The results also accord with the finding that the acoustic responses are locally linear in small distances~\cite{laufer2015study}. Meanwhile, linear interpolation has the largest parameter size here, which is $N(D+L)$ with $N$ being the size of the training examples. To show that the DNN model implicitly learns the RTF manifold structure, the target ILDs and the DNN generated ones are visualized in the low-dimensional space in Fig.~\ref{fig2:embed}. The manifolds resemble each other in the sense that the samples are clearly organized according to the poses.

\subsection{Performance on simulated data}

In the rest of the experiments, a more general setup is considered, e.g., in conference rooms, where the microphone positions are fixed and the source moves around in limited space. Following the previous setup, the receiver pose is fixed and the source pose has three degrees of freedom while HRTF is no longer used. The acoustic space is sampled uniformly as shown in Fig.~\ref{fig1:room}(b). Dense data samples are generated every 1 cm and there are in total 200$\times$100$\times$100 poses in the training set. 10,000 extra poses are randomly chosen for the evaluation.

The performance results are given in Table~\ref{table:dist}. The model is evaluated w.r.t. sample distances $\{$1, 2, 4, 8$\}$ cm in the training set and larger sample distance also means less training data. The prediction errors clearly go up with larger sample distances. This is more obvious for linear interpolation (using (\ref{eq:li}) with interpolation samples along the $z$ coordinate which reports lower errors than along the $x, y$ coordinates), the mean errors in ILD and IPD are more than double in the 2 cm case than that in the 1 cm case. DNN loses to linear interpolation again in small distances but slightly surpasses it starting from 4 cm. Note that the local linearity of the RTF manifold under the Euclidean distance measure~\cite{laufer2015study} holds within around 3.5 cm in this case. Here the performance of using $H_d(f)$, that is the prior knowledge of the RTF responses we can have from the free field assumption, to approximate the targets is given by: ILD, 3.53$\pm$0.053 and IPD, 0.62 $\pm$ 0.010.

\begin{table}[!h]
\caption{ILD/IPD prediction errors of DNN and linear interpolation w.r.t. sample distance. (mean$\pm$CI bound)}
\label{table:dist}
\begin{center}
\begin{tabular}{|l|c|c|c|c|}
  \hline
               & 1cm           & 2cm           & 4cm           & 8cm   \\ \hline \hline
   DNN-ILD    &3.01$\pm$.044 &3.03$\pm$.046  &3.10$\pm$.047 &3.23$\pm$.049   \\ \hline
   Linear-ILD &0.92$\pm$.017 &2.21$\pm$.036  &3.18$\pm$.049 &3.43$\pm$.052   \\ \hline
   DNN-IPD    &0.46$\pm$.008 &0.48$\pm$.009  &0.50$\pm$.009 &0.54$\pm$.009  \\ \hline
   Linear-IPD &0.17$\pm$.005 &0.38$\pm$.008  &0.53$\pm$.009 &0.57$\pm$.010  \\ \hline
\end{tabular}
\end{center}
\end{table}

\begin{figure}[!t]
\centering
\includegraphics[width=0.95\linewidth]{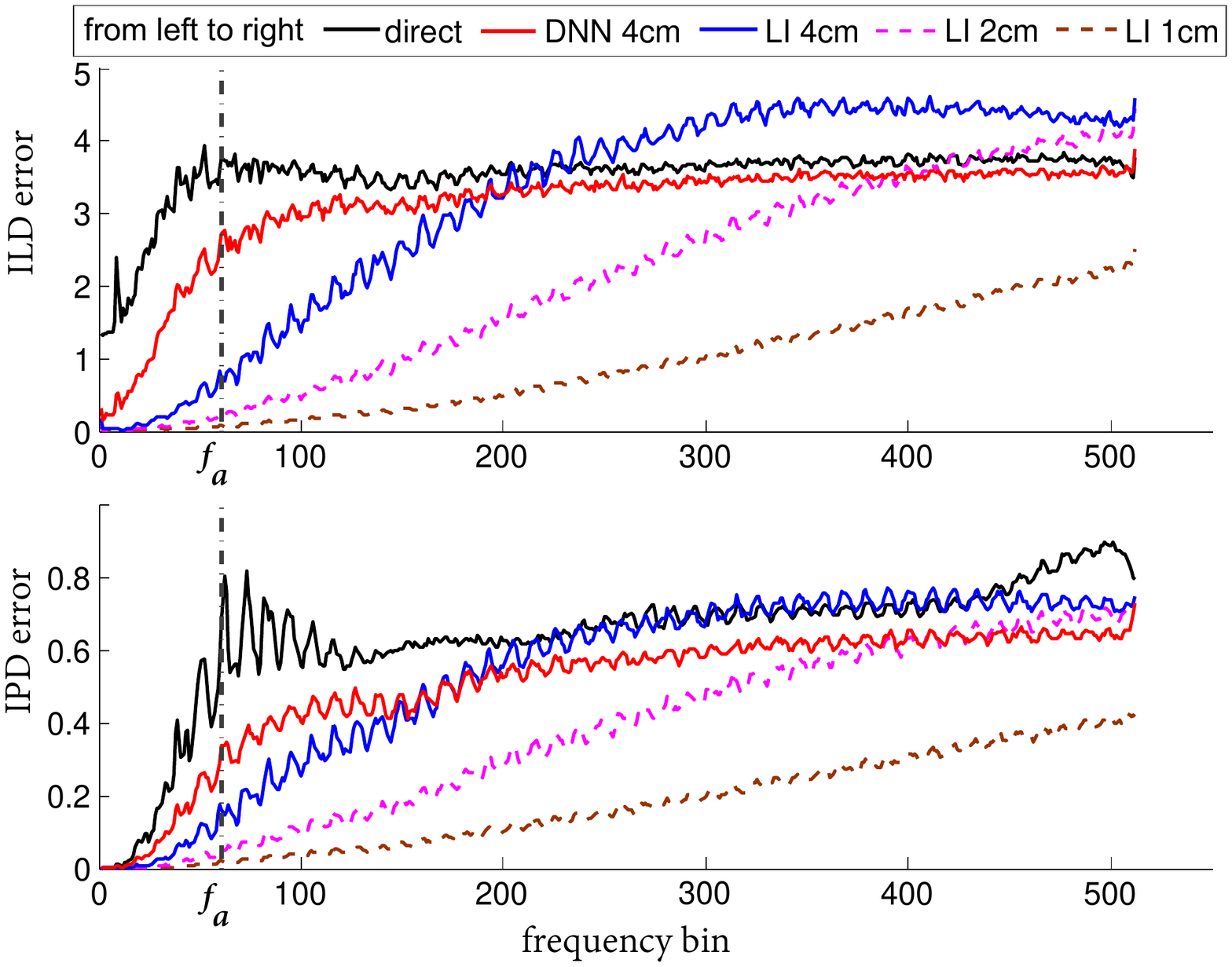}
\caption{The mean errors in ILD and IPD along the frequency axis. $f_a$ marks the spatial aliasing frequency.}
\label{fig:err-freq}
\end{figure}

The mean errors in different frequencies are plotted in Fig.~\ref{fig:err-freq}. The errors in the low frequencies are relatively smaller, especially below the spatial aliasing frequency $f_a$ as shown by $H_d(f)$ (direct) and the DNN model. In the high frequencies, the target values vary more rapidly on the manifold and the variations become harder to be captured by linearity. It is shown that DNN outperforms linear interpolation (LI) only in the high frequencies.

Considering the usage of white noise source signals in the RTF measurement process, there exist measurement errors: ILD 0.93$\pm$0.027, and IPD 0.14$\pm$0.004, which are given by running 2000 separate simulations for the same pose. The mean errors are close to that of the 1 cm case, which is one reason that no denser sampling is considered. The ambient noise is also considered as another factor and diffuse noise is manually added to the source signals in the training data at Signal-to-Noise Ratios (SNRs) $\{$30, 20, 10$\}$ dB. The results for the 2 cm sample distance are given in Table~\ref{table:noise}. The predictions errors differ slightly in the three cases, which means that both methods are quite robust to noise.

\begin{table}
\caption{ILD/IPD prediction errors w.r.t. SNR. (mean$\pm$CI bound)}
\label{table:noise}
\begin{center}
\begin{tabular}{|l|c|c|c|c|}
  \hline
                 & 30dB         & 20dB         & 10dB       \\ \hline \hline
   DNN-ILD     &3.03$\pm$.046  &3.03$\pm$.046 &3.05$\pm$.046  \\ \hline
   Linear-ILD  &2.31$\pm$.036  &2.32$\pm$.036 &2.36$\pm$.037   \\ \hline
   DNN-IPD     & 0.48$\pm$.009 & 0.48$\pm$.009 & 0.48$\pm$.009 \\ \hline
   Linear-IPD  & 0.40$\pm$.008 & 0.40$\pm$.008 &0.41$\pm$.008  \\ \hline
\end{tabular}
\end{center}
\end{table}


\section{Conclusion}

In this letter, we raised the RTF inverse regression problem for the first time and addressed it in the simplified stationary room environments. The trained DNN model directly generated the high-dimensional acoustic responses given the low-dimensional source poses. It performed better than the free field model and also captured the RTF manifold structure implicitly. The superior performance of linear interpolation in small distances supported the locality property of the RTFs. Experimental evaluations in a specific application or in a different test environment would be some valuable work to be addressed in the future.



%

\section*{Acknowledgment}

The authors would like to thank Antoine Deleforge and Sharon Gannot for their helpful discussions and comments.

  \newpage

\IEEEtriggeratref{15}


\bibliographystyle{IEEEtran}
\bibliography{bibliography}
%

\end{document}